\documentclass[prb,preprint,amsmath,showpacs,superscriptaddress]{revtex4}
\usepackage{amsmath}
\usepackage{graphicx}
\usepackage{bm}

\begin{document}

\title{Ultrafast extrinsic spin-Hall currents}

\author{E. Ya. Sherman}

\affiliation{Department of Physics and Institute for Optical Sciences, 
University of Toronto, 60 St. George Street, Toronto, Ontario, Canada M5S 1A7}

\author{Ali Najmaie}

\affiliation{Department of Physics and Institute for Optical Sciences, 
University of Toronto, 60 St. George Street, Toronto, Ontario, Canada M5S 1A7}

\author{H.M. van Driel }

\affiliation{Department of Physics and Institute for Optical Sciences, 
University of Toronto, 60 St. George Street, Toronto, Ontario, Canada M5S 1A7}

\author{Arthur L. Smirl }

\affiliation{Laboratory for Photonics and Quantum Electronics, 138 IATL, University of Iowa, Iowa City, Iowa 52242}

\author{J.E. Sipe}

\affiliation{Department of Physics and Institute for Optical Sciences, 
University of Toronto, 60 St. George Street, Toronto, Ontario, Canada M5S 1A7}

\begin{abstract}
We consider the possibility of ultrafast extrinsic spin-Hall currents,
generated by skew scattering following the optical injection of charge
or pure spin currents. We propose a phenomenological model for this
effect in quantum well structures. An injected charge current leads to a
spin-Hall-induced pure spin current, and an injected pure spin current leads
to a spin-Hall-induced charge current. The resulting spin or charge
accumulation can be measured optically.
\end{abstract}


\maketitle

\section{Introduction}

The spin-Hall effect (SHE), which leads to a spin current carried either by
electrons or holes driven out of the equilibrium, has attracted the
attention of both theorists and experimentalists because of its diverse and
interesting physics, and because of possible applications in spintronics.
Recent reviews of the current understanding of this effect have been
presented by Engel \textit{et al. }\cite{Engel06} and Schliemann \cite
{Schliemann06}. \ Usually two versions of the effect are distinguished. The 
\textit{extrinsic }spin-Hall current is due to a spin-dependent scattering
of electrons by the screened Coulomb potential of charged impurities \cite
{Dyakonov71a,Abakumov72,Hirsch99}, and arises from spin-orbit (SO) coupling
corrections to the scattering potential. The \textit{intrinsic} spin-Hall
current \cite{Engel06,Schliemann06,Sinova04} results from electron spin
precession due to SO coupling, and is understood as a band structure effect 
\cite{Murakami03}. In usual scenarios for observing both effects, the
electron system is initially close to the thermal equilibrium; a weak
applied electric field leads to electrical currents and, through various
mechanisms involving spin-orbit coupling, spin currents arise. Hence the
spin current arises as a response of the system to an external static or a
far-infrared electric field \cite{Rashba04}. The fact that the system is
close to equilibrium has at least two important consequences. First, the
initial electron distribution is known, and the well-established techniques
of response function calculations for a weakly perturbed systems can be
applied. Second, the random
potential of impurities, screened by the equilibrium distribution of
electrons \cite{Huang04}, is well-known and independent of the external
perturbation. At low temperatures, the charged impurities are screened on
the spatial scale of the Thomas-Fermi radius in the bulk, or of the quantum
well width for two-dimensional electrons.

Observations of intrinsic \cite{Wunderlich05} and extrinsic \cite
{Kato04,Sih05} spin-Hall effects have been reported recently, with the
existence and size of the effects extracted from experimentally determined
spin accumulation produced by the spin current. Thus the ability
to measure the spin accumulation is important for studies of
the spin-Hall effect. The spin accumulation pattern is strongly sensitive to
experimental conditions, and requires a thorough analysis for each
experiment and sample geometry \cite{Shytov04}.

Here we theoretically consider experiments involving the all-optical
generation of a strongly nonequilibrium extrinsic spin-Hall effect in
quantum wells. The approach is based on the coherent control of the
interband absorption of light, which allows charge and spin current to be
injected optically in bulk semiconductors and quantum wells \cite
{Bhat00,Najmaie03,Stevens02,Hubner03,Stevens03}. As a result, the system is
driven strongly out of equilibrium, with the injected electrons having
energies on the order of 150 meV above the bottom of the conduction
band, and velocities on the order of 1000 km/s; the injected 2D
electron areal concentrations $n$ are of the order of $10^{11}-10^{12}$ 
cm$^{-2}$. The injected charge and spin currents
then relax by collision of electrons with other carriers, phonons, and
impurities \cite{Rumyantsev04,Duc05} on the timescale of the order of $100$ $%
\mathrm{fs}$. Electron-hole, electron-electron, and electron-impurity
collisions can lead to a spin-Hall current via the extrinsic SHE. The spin
precession of photoexcited electrons due to Dresselhaus and Rashba-type SO
coupling can, in general, lead to the analog of intrinsic spin current. In this
paper we present a phenomenological model of the optically generated extrinsic
SHE; we comment briefly on the intrinsic effect in section 6.

Because the SHE we consider is all-optically generated, it can be studied
far from any bounding surfaces of a sample, and the kind of detailed
analysis of edge effects\ necessary in near-equilibrium experiments is not
required. Further, since in principle optical excitation with a range of
pulse widths can be considered, it should be possible to study the
timescales involved in the SHE in a controllable way, leading to a more
detailed understanding of the microscopic mechanisms responsible for it.
And the far-from-equilibrium nature of the
experimental scenarios we consider means that many assumptions implicitly
made in the study of near-equilibrium spin-Hall effects must be
reconsidered, and the subject seen in a much broader perspective.

We begin with a reminder of the experimental schemes for coherently
controlled injection of current and spin current.
We then consider the scattering and space charge effects that can be
important in optical experiments, estimate the size of the spin-Hall
effects, and finally summarize and discuss our results. While we focus on 
quantum well structures, many of our general conclusions will hold for experiments
on bulk samples as well. 

\begin{quotation}
\bigskip
\end{quotation}

\section{Injection scenarios}

Earlier we had suggested the use of spin currents generated all-optically
via infrared absorption \cite{Sherman05} or Raman scattering \cite
{Najmaie05a,Najmaie05b} to study spin Hall effects in doped quantum wells.
Here we consider utilizing the simpler mechanisms of current or spin current
injection via absorption across the band gap of an undoped quantum well. In
the experiments we consider, two pulses with carrier frequencies $\omega $
and $2\omega $ and polarizations $\mathbf{e}_{\omega }$ and $\mathbf{e}%
_{2\omega }$ are directed onto the QW, leading to an electric field 
\begin{eqnarray*}
\mathbf{E}(t) &=&\mathbf{e}_{\omega }E_{\omega }(t)e^{i\phi _{\omega
}}e^{-i\omega t}+\mathbf{e}_{2\omega }E_{2\omega }(t)e^{i\phi _{2\omega
}}e^{-2i\omega t} \\
&&+c.c.,
\end{eqnarray*}
where $E_{\omega }(t)$ and $E_{2\omega }(t)$ are slowly varying amplitudes.
An important parameter is the relative phase parameter $\Delta \phi =2\phi
_{\omega }-\phi _{2\omega }$, which is under the control of the
experimentalist. The photon energy $2\hbar \omega $ is above but close to
the bandgap $E_{g}$ of the quantum well, and interference of one- and
two-photon absorption occurs \cite{vanDriel01,Stevens04}. We take the
intensity of the incident pulses to vary in space as $\exp \left( -\rho
^{2}/\Lambda ^{2}\right) $, where $\rho =\sqrt{x^{2}+y^{2}}$ is the lateral
coordinate, and the spot size $\Lambda $ is of the order of few microns.
In the laser spot a relatively low-density electron-hole plasma is
generated, with possibly a net spin depending on the polarizations chosen
for the pulses \cite{Najmaie03}. Current and spin current can also be
injected, with the injection controlled by adjusting the polarization of the
two pulses \cite{Najmaie03,vanDriel01,Stevens04}. At a given $(x,y)$ the
injected areal electrical and spin current densities are given by $J^{i}$ and $%
K^{ij}$ respectively, where 
\begin{eqnarray*}
J^{i}/n &=&e\overline{v^{i}}, \\
K^{ij}/n &=&\frac{\hbar }{4}\overline{v^{i}\sigma ^{j}+\sigma ^{j}v^{i}},
\end{eqnarray*}
with the overbar denoting an average of the indicated quantity at $(x,y)$, $%
\mathbf{v}$ is the velocity, and $\hbar \mathbf{\sigma /}2$ the spin; $i$
and $j$ are Cartesian indices. We begin by restricting ourselves to the
electrons; we return to the holes at the end of section 5. Neglecting the
spin-splitting in the conduction miniband, we can write 
\begin{eqnarray*}
J^{i} &=&e\left( j_{+}^{i}+j_{-}^{i}\right) , \\
K^{iz} &=&\frac{\hbar }{2}\left( j_{+}^{i}-j_{-}^{i}\right) ,
\end{eqnarray*}
where we have introduced the (number) areal current density of spin up and
spin down electrons, $j_{+}^{i}$ and $j_{-}^{i}$ respectively, 
\[
j_{\pm }^{i}=\int \frac{d\mathbf{k}}{(2\pi )^{2}}v_{cc}^{i}(\mathbf{k}%
)f_{\pm }(\mathbf{k}), 
\]
where $\mathbf{v}_{cc}(\mathbf{k})$ is the diagonal velocity matrix element
in the conduction miniband at the two dimensional wavevector $\mathbf{k}%
=(k_{x},k_{y})$, and $f_{(+,-)}(\mathbf{k})d\mathbf{k}$ is the number of
spin up (down) electrons in the corresponding element of the phase space.
The resulting areal densities of spin-up and spin-down electrons are given
by 
\[
n_{\pm }=\int \frac{d\mathbf{k}}{(2\pi )^{2}}f_{\pm }(\mathbf{k}), 
\]
and we can then define average velocities for each spin projection at a
given $(x,y)$ as 
\[
v_{\pm }^{i}=j_{\pm }^{i}/n_{\pm }, 
\]
where $i$ $=x$ or $y$.

In this first analysis we consider ultrafast experiments performed at or
near room temperature, and a number of time scales can be identified. One is
the optical pulse length, which is on the order of $100$ $\mathrm{fs}.$
Another is that of momentum relaxation, which is typically also on the order
of $100$ $\mathrm{fs}$. We introduce appropriate relaxation times $\tau _{x}$
and $\tau _{y}$ below, associated with motion in the indicated directions,
and for the experiments we consider they can be different. But we generally
expect that they are both on the order of the momentum relaxation time.
Below we will derive a spin Hall scattering time $\tau _{\mathrm{sH}}$,
defined such that the average spin-Hall force on an electron of effective
mass $m^{*}$ with a given spin is 
\begin{equation}
\left\langle F_{\mathrm{sH}}\right\rangle =m^{*}\left\langle v\right\rangle
\tau _{\mathrm{sH}}^{-1},  \label{fspinhall}
\end{equation}
where $\left\langle v\right\rangle $ is the average of one of the $v_{\pm
}^{i}$ over the indicated cloud of spins, often referred to as the ''swarm''
velocity, and $\left\langle F_{\mathrm{sH}}\right\rangle $ is the corresponding
average of the spin-Hall force, which will be in a direction perpendicular
to $\left\langle v\right\rangle $. We will see below that $\tau _{\mathrm{sH}%
}\approx 10$ $\mathrm{ps}$ and, since the corresponding rates satisfy $%
1/\tau _{\mathrm{sH}}\gg 1/\tau _{i}$, the spin-Hall currents will be slaved
to the momentum relaxation. We will also see that space charge effects will
arise on a time scale larger than the $\tau _{i}$, leading to charge
oscillations that are overdamped. Finally, diffusion and recombination with
holes will become important on even longer time scales. These we neglect in
the treatment we present here, where our focus is on identifying the typical
spin-Hall displacements that can be expected within a few hundred
femtoseconds of the exciting optical pulses.

While the laser pulse is interacting with the semiconductor, spin or
electrical currents, or both, can be injected, depending on the excitation
geometry. Since the pulse width is comparable to the momentum relaxation
time, and since the momentum relaxation time itself will change as the
injected density increases during excitation, a full kinetic theory will
ultimately be required to trace the distributions $f_{\pm }(\mathbf{k})$
during excitation; any other approximation during the excitation phase is
inevitably a crude one. We make the simplest such approximation by
separating the excitation and transport regimes, beginning our transport
analysis after the spins are assumed to have been injected by the laser
pulses. The velocities $\left\langle v_{\pm }^{i}\right\rangle $ assumed
right after the injection, but before our transport analysis begins, can
then be taken immediately from the kind of simple Fermi's Golden Rule
calculations that have been presented in the literature \cite
{Bhat00,Najmaie03}, and at this level are independent of the density
injected; since the injection pulses are on the order of $100$ $\mathrm{fs}$%
, we naturally assume that the hole spins are completely relaxed at the
start of our transport analysis. Despite its simplicity, we believe that
more detailed calculations will confirm the estimates this prediction
provides of the early separation of the spin distributions.

We consider two excitation geometries that we label $(a)$ and $(b)$ (see
Fig. 1). In the first $(a)$ we take $\mathbf{e}_{\omega }$ to be oriented
along the $y$ direction and $\mathbf{e}_{2\omega }$ along the $x$ direction;
in the second $(b)$ we take both $\mathbf{e}_{\omega }$ and $\mathbf{e}%
_{2\omega }$ to be oriented along the $y$ direction. In $(a)$ we inject a
pure spin current if $\Delta \phi =0$ (initially $\left\langle
v_{+}^{y}\right\rangle =-\left\langle v_{-}^{y}\right\rangle >0$ and $%
\left\langle v_{\pm }^{x}\right\rangle =0$), and in $(b)$ a pure electrical
current if $\Delta \phi =\pi /2$ (initially $\left\langle
v_{+}^{y}\right\rangle =\left\langle v_{-}^{y}\right\rangle <0$ and $%
\left\langle v_{\pm }^{x}\right\rangle =0)$ \cite{vanDriel01}; we assume
these choices are made for the relative phase parameter. Then we can take
the velocities $\left\langle v_{\pm }^{i}\right\rangle $ to vary as 
\begin{eqnarray}
\frac{d}{dt}\left\langle v_{\pm }^{x}\right\rangle &=&C_{(a,b)\pm
}^{x}(t)+S_{(a,b)\pm }^{x}(t),  \label{dynamics} \\
\frac{d}{dt}\left\langle v_{\pm }^{y}\right\rangle &=&C_{(a,b)\pm
}^{y}(t)+S_{(a,b)\pm }^{y}(t),  \nonumber
\end{eqnarray}
where the vectors $\mathbf{C}_{a\pm }(t)$ and $\mathbf{S}_{a\pm }(t)$
describe the effects of space charge and scattering in geometry $(a)$, and
with $\mathbf{C}_{b\pm }(t)$ and $\mathbf{S}_{b\pm }(t)$ the corresponding
effects in geometry $(b)$.

\section{Scattering}

We turn first to the scattering terms, which in form are independent of the
excitation geometry, and can be written as 
\begin{equation}
\left[ 
\begin{array}{c}
S_{\pm }^{x} \\ 
S_{\pm }^{y}
\end{array}
\right] =\left[ 
\begin{array}{cc}
-\tau _{x}^{-1} & \mp \text{ }\tau _{\mathrm{sH}}^{-1} \\ 
\pm \text{ }\tau _{\mathrm{sH}}^{-1} & -\tau _{y}^{-1}
\end{array}
\right] \left[ 
\begin{array}{c}
\left\langle v_{\pm }^{x}\right\rangle \\ 
\left\langle v_{\pm }^{y}\right\rangle
\end{array}
\right] .  \label{scatt}
\end{equation}
We now confirm this form, and estimate $\tau _{\mathrm{sH}}$. To do this we
note that the number of injected holes will be equal to the number of
injected electrons, and for high excitation densities and initially clean, undoped
samples we can expect the scattering of the electrons from the holes to
dominate the scattering of the electrons from any impurities. We assume that
the heavy holes injected can be considered approximately fixed in
determining their effect on the electrons' motion. The Hamiltonian
describing the role spin-orbit coupling plays in the scattering process of
electron at ${\bm \rho }$ by a hole at ${\bm \rho }_{j}$ then takes the form 
\begin{equation}
H_{\mathrm{sH}}\left( {\bm \rho }-{\bm \rho }_{j}\right) =\mp \frac{\hbar
\gamma }{4m^{*}E_{g}}\left[ \nabla _{\bm \rho }U\left( \left| {\bm \rho }-{%
\bm \rho }_{j}\right| \right) ,\mathbf{p}\right] ,\qquad  \label{spinorbit}
\end{equation}
where the upper (lower) sign refers to a spin up (down) electron, $\gamma
\equiv 2\Delta /E_{g},$ where $\Delta $ is the energy difference between the
split-off and valence band, and the potential energy $U$ describes Coulomb
interaction between electron and hole confined in the quantum well$.$ The
total spin-orbit force acting on an electron is therefore:

\begin{equation}
\mathbf{F}_{\mathrm{sH}}\left( {\bm \rho }\right) =\sum_{j}\mathbf{f}_{%
\mathrm{sH}}({\bm \rho} -{\bm \rho}_{j}),  \label{forcefield}
\end{equation}
where the force $\mathbf{f}_{\mathrm{sH}}({\bm \rho -\rho }_{j})$ on the
electron due to hole $j$ is given by 
\begin{equation}
\mathbf{f}_{\mathrm{sH}}({\bm \rho} -{\bm \rho}_{j})=-\nabla _{{\bm \rho }%
}H_{\mathrm{sH}}({\bm \rho} -{\bm \rho}_{j}).  \label{holeforce}
\end{equation}
The mean value of $\mathbf{F}_{\mathrm{sH}}$, which we seek, is the spin-
and velocity-dependent electromotive force acting on electrons. The
long-range potential energy of electron-hole interaction $U(\rho)$,

\[
U\left( \rho \right) =\frac{2e^{2}}{w\epsilon }\int_{0}^{w}\sin ^{2}\left(
\pi \frac{z_{1}}{w}\right)\int_{0}^{w}\sin ^{2}\left( \pi \frac{z_{2}%
}{w}\right) \frac{dz_{2} dz_{1}}{\sqrt{\rho ^{2}+\left( z_{1}-z_{2}\right) ^{2}}}%
,\qquad 
\]
where $w$ is the width of the quantum well and $\epsilon $ the background
dielectric constant, exhibits a weak logarithmic divergence at small
distances 
\[
U\left( \rho \right) \approx \left\{ 
\begin{array}{ll}
3\left( e^{2}/\epsilon w\right) \ln \left( w/\rho \right) , & \rho \ll w, \\ 
e^{2}/\epsilon \rho , & \rho \gg w.
\end{array}
\right. \qquad 
\]
Due to the high energy of electrons, their relatively low density, \ and the
short times of interest here, screening effects can be assumed to be small
and therefore have been neglected.

We calculate the skew scattering in a classical approximation, considering
an electron with momentum $\mathbf{p}=\left( 0,p\right) $ moving in the
force field (\ref{forcefield}). For a spin-up electron at the origin, the $x$%
-component of the force on it due to a hole at a distance $\rho $ in the $xy$
plane from the electron, making an angle $\zeta $ from the $x$-axis (see
Fig. 2) is 
\begin{eqnarray*}
f_{\mathrm{sH}}^{x}({\bm \rho }-{\bm \rho }_{h})|_{{\bm \rho =0}} &=&\frac{%
\hbar \gamma }{4m^{*}E_{g}}p\left( \frac{\partial ^{2}U(\left| {\bm \rho }-{%
\bm \rho }_{h}\right| )}{\partial x^{2}}\right) _{\boldsymbol{\rho =0}} \\
&=&-\frac{\hbar \gamma }{4m^{*}E_{g}}p\left( \cos ^{2}\zeta \frac{\partial
^{2}U(\rho _{h})}{\partial \rho _{h}^{2}}+\frac{\sin ^{2}\zeta }{\rho _{h}}%
\frac{\partial U(\rho _{h})}{\partial \rho _{h}}\right)  \\
&\equiv &g(\rho _{h},\zeta ,p)
\end{eqnarray*}
where ${\bm \rho }_{h}=(\rho _{h}\cos \zeta ,\rho _{h}\sin \zeta )$. The
mean value of this force, for holes uniformly distributed with an areal
density $n$, is given by 
\begin{eqnarray}
\left\langle F_{\mathrm{sH}}^{x}\right\rangle  &=&n\int_{0^{+}}^{\infty
}d\rho _{h}\rho _{h}\int_{\zeta =0}^{2\pi }g(\rho _{h},\zeta ,p)d\zeta \text{
\ \ \ \ \ \ }  \label{forceaverage} \\
&=&-3\pi \frac{e^{2}n}{\epsilon }\frac{p}{\hbar k_{g}}\frac{1}{k_{g}w}, 
\nonumber
\end{eqnarray}
where the integral over $\rho _{h}$ is done for a small lower bound, which
is then allowed to approach zero at the end of the calculation, and 
$k_g=2\sqrt{E_{g}m^{*}/\gamma\hbar^2}$ is the  wavevector associated with the
spin-orbit coupling strength (4). Now
associating $p$ with the $m^{*}\left\langle v\right\rangle $ of (\ref
{fspinhall}), we can identify 
\begin{equation}
\tau _{\mathrm{sH}}=\frac{\hbar k_{g}^{2}w\epsilon }{3\pi ne^{2}},
\label{tausH}
\end{equation}
and, repeating the calculation for a spin-down electron, confirm the
signs in (\ref{scatt}). For GaAs quantum wells we expect typical values of $%
k_{g}^{2}=2\times 10^{15}$ $\mathrm{cm}^{-2}$ and $w=10$ $\mathrm{nm}$; for
injected areal densities of $n=10^{12}$ $\mathrm{cm}^{-2}$, we find $\tau_{\mathrm{sH}}\approx $ $10$ $\mathrm{ps.}$ This is in comparable with a
result of 20 ps found earlier \cite{Hankiewicz06} for the skew-scattering
time for Coulomb drag effects, despite the different nature of the
scattering potential. From (\ref{forceaverage}) we see that the spin-orbit
force is small in comparison to the characteristic Coulomb force $%
ne^{2}/\epsilon $, as expected. 

This treatment of the spin-Hall force is clearly a simple one; besides its
mean field nature, the so-called ''side-jump'' component of the scattering
is neglected. Our main use here of the result (\ref{tausH}) is to help us in
our estimates below of the order of the magnitude of the size of the charge
and spin displacements that will result. \ In that respect the neglect of
the side-jump component, which typically does not affect the order of
magnitude of the spin-Hall scattering in the region of large momenta, should
not lead to serious error \cite{Huang04,Engel05}.

To touch base with the traditional literature on the spin-Hall effect in
near-equilibrium systems, we can use our phenomenological scattering
description (\ref{scatt}), with an assumed static electric field in the $y$
direction on a sample of uniform density, 
\[
\frac{d}{dt}\left[ 
\begin{array}{c}
v_{\pm }^{x} \\ 
v_{\pm }^{y}
\end{array}
\right] =\left[ 
\begin{array}{cc}
-\tau ^{-1} & \mp \text{ }\tau _{\mathrm{sH}}^{-1} \\ 
\pm \text{ }\tau _{\mathrm{sH}}^{-1} & -\tau ^{-1}
\end{array}
\right] \left[ 
\begin{array}{c}
v_{\pm }^{x} \\ 
v_{\pm }^{y}
\end{array}
\right] +\left[ 
\begin{array}{c}
0 \\ 
eE^{y}/m^{*}
\end{array}
\right] , 
\]
to make a steady state calculation; here, for simplicity, we have put $\tau
_{x}=\tau _{y}=\tau $. We find an areal current density $J^{y}=\sigma E^{y}$, where since $%
\tau /\tau _{\mathrm{sH}}\ll 1$ the conductivity $\sigma =(e^{2}n\tau
/m^{*})(1+\tau ^{2}/\tau _{\mathrm{sH}}^{2})^{-1}$ is only slightly modified
from the usual result $e^{2}n\tau /m^{*}$, and an areal spin current density 
$K^{xz}=\sigma _{\mathrm{sH}}E^{y}$, where 
\begin{equation}
\sigma _{\mathrm{sH}}=\frac{\hbar }{2\left| e\right| }\frac{\tau }{\tau _{%
\mathrm{sH}}}\sigma .  \label{sigmasH}
\end{equation}
The ratio $\tau /\tau _{\mathrm{sH}}\approx 0.01$ we find here is
considerably less than predicted for 2DEG systems \cite{Hankiewicz06},
because the relaxation time $\tau $ is much longer in those samples. But our
result is comparable to the spin-Hall angle $\tau /\tau _{sH}$ predicted for
doped quantum wells with impurities in the well \cite{Huang04}, a situation
roughly comparable to ours, but with impurities playing the role of the
holes.

\section{Space charge effects}

In the experimental geometries we consider, space charge effects arise as
the holes and electrons separate; their strength can be characterized by the
plasma frequency. Consider first a simple example where the center of charge
of an electron distribution is displaced a distance $x$ from the center of
charge of the corresponding hole distribution, assumed fixed. Neglecting any
diffusion of the charges and considering only ballistic motion, familiar
elementary arguments give that the velocity $v$ of the center of the
electron charge distribution satisfies the equation 
\[
\frac{dv}{dt}=-\Omega ^{2}x,
\]
where $\Omega $ is the plasma frequency for the given charge pattern. For
a rigid Gaussian distribution of electrons and holes 
of the form $n\exp\left(-\rho ^{2}/\Lambda ^{2}\right)$ in a single quantum well \cite
{Ando82}, the effective plasma frequency is given by 
\begin{equation}
\Omega ^{2}=\left( \frac{\pi }{2}\right) ^{3/2}\frac{ne^{2}}{\epsilon
m^{*}\Lambda }.  \label{plasma}
\end{equation}
This dependence of the plasma frequency on $\Lambda $ is characteristic of 2D
plasma oscillations, and does not arise for plasma oscillation in three
dimensions; in the geometries we consider here we can take $\Lambda $ to be
of the order of the laser spot size injecting the carriers. 
For characteristic GaAs
parameters, an assumed injected areal carrier density of $n=10^{12}$ cm$^{-2}$,
and $\Lambda =2$ $\mu\mathrm{m}$, we find $\Omega^{2}\tau^{2}\approx 0.03$, where we
have taken a momentum relaxation time of $\tau =100$ fs. In a multiple
quantum well (MQW) structure consisting of $N_{\rm QW}$ identical quantum wells
each with areal density $n$, and far from any dielectric/air interface, 
the plasma frequency is approximately given by 
$\Omega^{2}=N_{\rm QW}(\pi /2)^{3/2}ne^{2}/(\epsilon m^{\ast}\Lambda)$, if
the total thickness of the structure is much less than $\Lambda $; this 
condition is well-satisfied for widely used structures with $N_{\rm QW}\lesssim
10$ and the sum of the well and barrier thicknesses $\lesssim 30$ nm. 
If all the quantum wells are much closer than $\Lambda$ to a dielectric/air 
interface, $\Omega^{2}$ will be enhanced  by a factor of $2\epsilon/(\epsilon+1)$. Even
for such structures we have $\Omega ^{2}\tau^{2}<1$, and momentum
relaxation will control the evolution of carrier velocities initially, space
charge effects arising only after the injected velocities have considerably
slowed.  Nonetheless, since $%
\Omega ^{-1}$ is much shorter than typical diffusion times, we can estimate
the consequences of space charge effects by neglecting any change in the
distribution of electrons other than the motion of their center of charge $%
\left\langle {\bm \rho }_{\pm }\right\rangle =\left( \left\langle x_{\pm
}\right\rangle ,\left\langle y_{\pm }\right\rangle \right) $, where $%
\left\langle x_{+}\right\rangle $ is the $x$-component of the center of
charge of spin-up electrons, etc.

Charge separation occurs in characteristically different ways in our two
geometries (see Fig. 1). \ In geometry $(a)$ the two electron
distributions (spin-up and spin-down) initially separate from a hole
distribution remaining, in our neglect of hole velocities, centered at the
origin. A "quadrupole-type" charge separation in the $y$ direction results,
and space charge effects will here be small. The charge separation of
spin-up and spin-down carriers in the $x$ direction will be the same
direction, however, and there will be none of the cancellation that occurs
in the $y$ direction. Hence, to first approximation, in geometry $(a)$ we
can write 
\begin{equation}
\left[ 
\begin{array}{c}
C_{a\pm }^{x} \\ 
C_{a\pm }^{y}
\end{array}
\right] =\left[ 
\begin{array}{cc}
-\Omega _{x}^{2} & 0 \\ 
0 & 0
\end{array}
\right] \left[ 
\begin{array}{c}
\left\langle x_{\pm }\right\rangle \\ 
\left\langle y_{\pm }\right\rangle
\end{array}
\right] ,  \label{spacea}
\end{equation}
where we will take $\Omega _{x}$ (and $\Omega _{y}$ below) to be given by (%
\ref{plasma}) or its generalization to multiple 
quantum wells, but use the subscript to indicate the geometry. In geometry $%
(b)$ the situation is reversed; there are significant space charge effects
only in the $y$ direction, and we have 
\begin{equation}
\left[ 
\begin{array}{c}
C_{b\pm }^{x} \\ 
C_{b\pm }^{y}
\end{array}
\right] =\left[ 
\begin{array}{cc}
0 & 0 \\ 
0 & -\Omega _{y}^{2}
\end{array}
\right] \left[ 
\begin{array}{c}
\left\langle x_{\pm }\right\rangle \\ 
\left\langle y_{\pm }\right\rangle
\end{array}
\right] .  \label{spaceb}
\end{equation}

\section{Charge and spin dynamics}

We can now assemble and solve our approximate dynamical equations for the
two geometries we have considered. We begin with geometry (a). Here the
injection yields $\left\langle v_{\pm }^{y}(0)\right\rangle \equiv \pm v_{0}$%
, where $v_{0}$ is the initial speed of the injected spins, and $%
\left\langle v_{\pm }^{x}(0)\right\rangle =0$. Combining the scattering (\ref
{scatt}) and space charge (\ref{spacea}) effects in our dynamical equations (%
\ref{dynamics}), we find 
\begin{equation}
\frac{d}{dt}\left[ 
\begin{array}{c}
\left\langle v_{\pm }^{x}\right\rangle  \\ 
\left\langle v_{\pm }^{y}\right\rangle 
\end{array}
\right] =\left[ 
\begin{array}{cc}
-\tau _{x}^{-1} & \mp \text{ }\tau _{\mathrm{sH}}^{-1} \\ 
\pm \text{ }\tau _{\mathrm{sH}}^{-1} & -\tau _{y}^{-1}
\end{array}
\right] \left[ 
\begin{array}{c}
\left\langle v_{\pm }^{x}\right\rangle  \\ 
\left\langle v_{\pm }^{y}\right\rangle 
\end{array}
\right] +\left[ 
\begin{array}{cc}
-\Omega _{x}^{2} & 0 \\ 
0 & 0
\end{array}
\right] \left[ 
\begin{array}{c}
\left\langle x_{\pm }\right\rangle  \\ 
\left\langle y_{\pm }\right\rangle 
\end{array}
\right] .  \label{dynamicsa}
\end{equation}
In solving for $\left\langle v_{\pm }^{y}\right\rangle $ we can justifiably
neglect the back-effect due to the spin-Hall scattering from the small $%
\left\langle v_{\pm }^{x}\right\rangle ,$ which is itself generated from
spin-Hall scattering. We find 
\[
\left\langle v_{\pm }^{y}\right\rangle =\pm v_{0}e^{-t/\tau _{y}},
\]
and using this in (\ref{dynamicsa}) we can solve for $\left\langle v_{\pm
}^{x}\right\rangle $, together $\left\langle y_{\pm }\right\rangle $ and $%
\left\langle x_{\pm }\right\rangle $, using $d\left\langle y_{\pm
}\right\rangle /dt=\left\langle v_{\pm }^{y}\right\rangle $ and $%
d\left\langle x_{\pm }\right\rangle /dt=\left\langle v_{\pm
}^{x}\right\rangle $. The results are 
\[
\left\langle y_{\pm }\right\rangle =\pm v_{0}\tau _{y}\left( 1-e^{-t/\tau
_{y}}\right) 
\]
and 
\begin{eqnarray*}
&&\left\langle x_{\pm }\right\rangle  \\
&=&\mp a_{\mathrm{PSC}}\left\{ e^{-t/\tau _{y}}+\left( \frac{\tau _{x}^{-}}{%
\tau _{x}^{+}-\tau _{x}^{-}}\left( \frac{\tau _{x}^{+}}{\tau _{y}}-1\right)
-1\right) e^{-t/\tau _{x}^{+}}\right.  \\
&&\hspace{1cm}+\left. \frac{\tau _{x}^{-}}{\tau _{x}^{+}-\tau _{x}^{-}}%
\left( 1-\frac{\tau _{x}^{+}}{\tau _{y}}\right) e^{-t/\tau _{x}^{-}}\right\}
,
\end{eqnarray*}
where 
\[
a_{\mathrm{PSC}}=\frac{v_{0}}{\tau _{\mathrm{sH}}}\frac{\tau _{y}^{2}\tau
_{x}}{\tau _{x}-\tau _{y}+\Omega _{x}^{2}\tau _{y}^{2}\tau _{x}},
\]
and 
\begin{equation}
\frac{1}{\tau _{x}^{\pm }}\equiv \frac{1}{2}\left( \frac{1}{\tau _{x}}\pm 
\sqrt{\frac{1}{\tau _{x}^{2}}-4\Omega _{x}^{2}}\right) ;  \label{taupm}
\end{equation}
note that $\tau _{x}^{+}\approx \tau _{x}^{-1}$ and sets a short time scale
for the evolution of $\left\langle x_{\pm }\right\rangle $, and $\tau
_{x}^{-}\approx (\Omega _{x}^{2}\tau _{x})^{-1}$, which sets a longer time
scale. The results for $\left\langle y\right\rangle \equiv \left|
\left\langle y_{\pm }\right\rangle \right| $ and $\left\langle
x\right\rangle \equiv \left| \left\langle x_{\pm }\right\rangle \right| $
are given in Fig. 3a. We have assumed that electrons are injected about $%
150$ $\mathrm{meV}$ above the bottom of the conduction band, leading to an
injected swarm velocity of $v_{0}=500$ km/s for
the electrons. In choosing the other parameters we have in mind a typical
momentum relaxation time $\tau =100$ $\mathrm{fs}$. Thus we set $\tau
_{x}=100$ $\mathrm{fs}$ but take $\tau _{y}=50$ $\mathrm{fs}$, shorter than
the momentum relaxation time, because the spin current can be expected to
relax on a faster time scale than current, there being many scattering
mechanisms that can redistribute spin without redistributing momentum \cite
{Rumyantsev04,Duc05}; in line with the discussion in section 4 above and using 
Eq.(10) to evaluate  the plasma frequency, we
take $\Omega_{x}\tau=0.15$ (weak space charge effects, solid lines) 
for a single quantum well and $\Omega_{x}\tau=0.4$ (considerable space charge effects, dashed lines)
for the $N_{\rm QW}=8$ structure, with the wells 
far from a dielectric/air interface. As a result of space charge effect, 
$\left\langle x\right\rangle $ decreases back to zero at the time scale dependent 
on $\Omega _{x}\tau$ as the dipole
charge separation pulls the electrons back to the holes. 
Since we have neglected the weaker, quadrupole
type space charge effects that arise in the $y$ direction, even here $%
\left\langle y\right\rangle $ does not relax at long times. A more
realistic calculation of space charge effects would of course show a
decrease in $\left\langle y\right\rangle $, but on a longer time scale than
that  exhibited by the decay of $\left\langle x\right\rangle $ in 
Fig. 3a.

For geometry (b) our dynamical equations are 
\begin{equation}
\frac{d}{dt}\left[ 
\begin{array}{c}
\left\langle v_{\pm }^{x}\right\rangle  \\ 
\left\langle v_{\pm }^{y}\right\rangle 
\end{array}
\right] =\left[ 
\begin{array}{cc}
-\tau _{x}^{-1} & \mp \text{ }\tau _{\mathrm{sH}}^{-1} \\ 
\pm \text{ }\tau _{\mathrm{sH}}^{-1} & -\tau _{y}^{-1}
\end{array}
\right] \left[ 
\begin{array}{c}
\left\langle v_{\pm }^{x}\right\rangle  \\ 
\left\langle v_{\pm }^{y}\right\rangle 
\end{array}
\right] +\left[ 
\begin{array}{cc}
0 & 0 \\ 
0 & -\Omega _{y}^{2}
\end{array}
\right] \left[ 
\begin{array}{c}
\left\langle x_{\pm }\right\rangle  \\ 
\left\langle y_{\pm }\right\rangle 
\end{array}
\right] ,  \label{dynamicsb}
\end{equation}
and space charge effects enter immediately in the evolution of $\left\langle
v_{\pm }^{y}\right\rangle ,$ since we are injecting a net current with $%
\left\langle v_{\pm }^{y}(0)\right\rangle \equiv -v_{0}$. Again we have $%
\left\langle v_{\pm }^{x}(0)\right\rangle =0$, so we neglect the small
spin-Hall contribution of the generated $\left\langle v_{\pm
}^{x}\right\rangle $ on $\left\langle v_{\pm }^{y}\right\rangle $ and,
solving the second of (\ref{dynamicsb}) together with $d\left\langle y_{\pm
}\right\rangle /dt=\left\langle v_{\pm }^{y}\right\rangle $ yields 
\[
\left\langle v_{\pm }^{y}\right\rangle =\frac{-v_{0}}{\sqrt{\tau
_{y}^{-2}-4\Omega _{y}^{2}}}\left\{ \frac{1}{\tau _{y}^{+}}\exp \left[
-t/\tau _{y}^{+}\right] -\frac{1}{\tau _{y}^{-}}\exp \left[ -t/\tau
_{y}^{-}\right] \right\} ,
\]
and 
\[
\left\langle y_{\pm }\right\rangle =\frac{-v_{0}}{\sqrt{\tau
_{y}^{-2}-4\Omega _{y}^{2}}}\left\{ \left( 1-\exp \left[ -t/\tau
_{y}^{+}\right] \right) -\left( 1-\exp \left[ -t/\tau _{y}^{-}\right]
\right) \right\} 
\]
where $\tau _{y}^{\pm }$ are defined in terms of $\tau _{y}$ and $\Omega _{y}
$ as are $\tau _{x}^{\pm }$ in terms of $\tau _{x}$ and $\Omega _{x}$ (\ref
{taupm}). The minus sign in front of $v_{0}$ here is due to the fact that at 
$2\phi _{\omega }-\phi _{2\omega }=\pi /2,$ which maximizes the injected
current, the electrons move in the $-y-$direction, as shown in Fig.1(b).
A straightforward integration of the second of (\ref{dynamicsb}) and $%
d\left\langle x_{\pm }\right\rangle /dt=\left\langle v_{\pm
}^{x}\right\rangle $ then yields 
\begin{eqnarray*}
&&\left\langle x_{\pm }\right\rangle  \\
&=&\mp a_{C}\left\{ \frac{\tau _{y}^{+}}{\tau _{x}-\tau _{y}^{+}}e^{-\tau
/\tau _{y}^{+}}-\left( \frac{\tau _{y}^{+}}{\tau _{x}-\tau _{y}^{+}}-\frac{%
\tau _{y}^{-}}{\tau _{x}-\tau _{y}^{-}}\right) e^{-t/\tau _{x}}-\frac{\tau
_{y}^{-}}{\tau _{x}-\tau _{y}^{-}}e^{-t/\tau _{y}^{-}}\right\} ,
\end{eqnarray*}
where 
\[
a_{C}=\frac{-v_{0}}{\tau _{\mathrm{sH}}}\frac{\tau _{x}}{\sqrt{\tau
_{y}^{-2}-4\Omega _{y}^{2}}}.
\]
The results for $\left\langle y\right\rangle \equiv \left| \left\langle
y_{\pm }\right\rangle \right| $ and $\left\langle x\right\rangle \equiv
\left| \left\langle x_{\pm }\right\rangle \right| $ are plotted in Fig.
Here we take $\tau _{x}=50$ $\mathrm{fs}$
and $\tau _{y}=100$ $\mathrm{fs}$, since in this geometry the form
corresponds to the relaxation of spin current and the latter the relaxation
of current; all other parameter values are as used above.
In this excitation geometry the space charge effects impact the evolution of $\left\langle
x\right\rangle $ indirectly within our model, through their effects on $%
\left\langle v_{\pm }^{y}\right\rangle $ as the electrons pull away from the
holes in the $y$ direction; the resulting decrease in $\left\langle v_{\pm
}^{y}\right\rangle $ leads to less spin-Hall current than would otherwise be
induced. Since here the separation of $\left\langle x_{\pm }\right\rangle $
leads to a quadrupole type space charge effect that we neglect, there is no
direct space charge effect on $\left\langle x\right\rangle $ in our model.

Despite the differences in the way space charge effects modify the dynamics
in these two geometries, the inequalities $\Omega _{x,y}^{2}\tau
_{x,y}^{2}\ll 1$ guarantee that these effects enter on a longer time scale
than the current and spin current relaxation times ($\tau _{y}$ and $\tau
_{x}$ respectively in our scenarios), and thus the maximum values of $%
\left\langle y\right\rangle $ and $\left\langle x\right\rangle $ can be
estimated neglecting space charge effects; these estimates are 
\begin{eqnarray}
\left\langle y\right\rangle _{\max } &=&v_{0}\tau _{y},  \label{disresults}
\\
\left\langle x\right\rangle _{\max } &=&v_{0}\tau _{y}\frac{\tau _{x}}{\tau
_{\mathrm{sH}}}.  \nonumber
\end{eqnarray}
Comparing the dashed and solid lines in Figs. 2a,b shows that within our
model these estimates are valid to within about 20\%, and somewhat better
than that for the ratio $\left\langle x\right\rangle _{\max }/\left\langle
y\right\rangle _{\max }$. Note that this ratio will be different in geometry 
$(a)$ than in geometry $(b)$, because in the former $\tau _{x}$ identifies
the current relaxation time, and in the latter the spin current relaxation
time. However, these times should be the same within a factor of two \cite
{Rumyantsev04,Duc05}, which is what we have assumed above. For the
parameters we have adopted here, in geometry $(a)$ we find $\left\langle
y\right\rangle _{\max }=25$ $\mathrm{nm}$ and $\left\langle x\right\rangle
_{\max }=0.25$ $\mathrm{nm}$; in geometry $(b)$ we find $\left\langle
y\right\rangle_{\max }=50$ $\mathrm{nm}$ and $\left\langle x\right\rangle
_{\max }=0.25$ $\mathrm{nm}$.

We conclude this section by returning to the holes, the motion of which we
have neglected in our analysis. In the approximation of parabolic bands, the
injected momentum of the holes will be equal in magnitude to the injected
momentum of the electrons, and in the opposite direction. Even if we move
beyond this approximation, we can generally expect the swarm velocity of the
holes to be much less than that of the electrons, due to their larger
effective mass. Therefore, the holes will move considerably less than the
electrons, and their motion will not qualitatively affect our results. This
is because what is crucial to our dynamical equations is the separation of the
electrons from the holes. And to first approximation that can be described
by the equations used here, with at most a modification of the electron
effective mass to describe the motion of the relative electron-hole
separation. For longer time scales, of course, a more detailed dynamical
treatment of the electron-hole plasma would be necessary.

\section{Summary and Discussion}

We have presented a simple theoretical description of what might be called
an "ultrafast spin-Hall effect," or rather a family of such effects. In one
scenario a spin current is optically injected in the sample, and the
spin-Hall effect leads to an electrical current; in the other scenario an
electrical current is optically injected in the sample, and the spin-Hall
effect leads to a pure spin current. Although our analysis has focused on 
updoped quantum wells, it should hold qualitatively for intrinsic bulk
samples as well, with appropriate values 
for plasma frequency and relaxation times.
And while we have focused on optical injection via an
interference between one- and two-photon absorption across the band gap \cite
{Bhat00,Najmaie03,vanDriel01,Stevens04}, other injection scenarios involving
the absorption of a single laser pulse \cite{Sherman05,Bhat05,Zhao05}, or
Raman scattering in the infrared \cite{Najmaie05a} or visible \cite
{Najmaie05b} should also lead to such ultrafast spin-Hall effects; we plan
to turn to these in later communications. The ultrafast nature of these
effects arises because the currents (or spin currents) can be injected on
the timescale of $100$ $\mathrm{fs}$, and the resulting spin currents (or
currents) that result from the spin-Hall effect are slaved to the injection
and relaxation of these injected currents, which also occurs on a timescale
of $100$ $\mathrm{fs}$. Thus they should be distinguished from other effects
that follow optical injection but occur on a longer time scale, such as the
proposal of Bakun \textit{et al.} \cite{Bakun85}, where the spin current
arises due a relatively slow ambipolar diffusion of injected spin-polarized
electrons with their spins precessing in an applied magnetic field.

The treatment we have presented here is very elementary, most importantly in
that we have artificially separated the injection and relaxation processes.
In practice these occur simultaneously, and a full kinetic treatment will be
essential to give a correct description of the injection and relaxation
processes; such a calculation is underway. Our treatment of space
charge effects is also a very simple one. But for the subpicosecond time
scales of interest, and typical spot sizes, we have argued that for a single
quantum well, or a MQW sample with a few wells, the main consequence of
space charge effects will be to relax the maximum spin and charge
displacements that are generated. Those maximum displacements can be
reasonably estimated neglecting space charge effects. Indeed, in quantum
well geometries the space charge effects can be reduced to some extent by
using a larger spot size than is usual, and hence decreasing $\Omega$
(10). In the geometry $(a)$ we considered, the charge displacement $%
\left\langle x\right\rangle _{\max }$ could be observed by the change in
light transmission through the quantum wells, while in geometry $(b)$ the
spin displacement $\left\langle x\right\rangle _{\max }$ could be observed
by monitoring the absorption of a circularly polarized probe pulse. The
distances involved are on the order of those observed in other experiments 
\cite{Zhao05}, and so experimental study of these ultrafast effects should
be feasible. Preliminary observations of these effects have already been 
reported \cite{Zhao06}, however in experiments that have not time-resolved 
the motion of the carriers.

The moniker ''ultrafast spin-Hall effects'' is truly justified for these
phenomena because the ratio of the skew-scattering-induced effect (whether
it be current or spin current) to the directly injected effect (spin current
or current, respectively), is given by 
\[
\frac{\left\langle x\right\rangle _{\max }}{\left\langle y\right\rangle
_{\max }}=\frac{\tau _{x}}{\tau _{\mathrm{sH}}}
\]
(see (\ref{disresults})) which, except for the subtlety mentioned after 
(\ref{disresults}), is essentially the same factor determining the ratio of the
spin Hall conductivity to the conductivity, $\sigma _{\mathrm{sH}}/\sigma $
(see (\ref{sigmasH})). Hence a measurement of $\left\langle x\right\rangle
_{\max }/\left\langle y\right\rangle _{\max }$ in these ultrafast
experiments is essentially a direct analog of the measurement of the
spin-Hall voltage in more usual, near-equilibrium transport experiments. 

Of course, this simple analogy can break down in MQW experiments using
samples with a large number of wells. The significance of space charge
effects is characterized by the product $\Omega ^{2}\tau ^{2}$, which is
proportional to the number of quantum wells in a sample. As we consider
increasing $\Omega^{2}\tau^{2}$, the first effect is a small, benign
decrease in the maximum currents and spin currents, as shown in Fig. 3.  But
as $\Omega ^{2}\tau ^{2}$ approaches unity, a clear separation between the
fast increase and slow relaxation times for the currents becomes impossible,
and the overdamped motion of the electron cloud is replaced by damped
oscillations. Since in practice this will happen as $\Omega ^{2}\tau _{p}^{2}
$ also approaches unity, where $\tau _{p}$ is the duration of the excitation
pulses, in this regime injection, scattering, and space charge effects will
all have to be considered together in a quantum kinetic description. \ We
also mention that both $\tau $ and $\tau_{\rm sH}$ will be modified as we move
to samples with larger numbers of quantum wells, since electrons in one well
will interact to some extent with holes in all wells. Hence, just as $\Omega 
$ is modified by moving to samples of larger numbers of quantum wells, $\tau 
$ and $\tau _{\rm sH}$ will be modified as well.

But while added complexity can arise due to space charge effects in MQW
samples, qualitatively new physics will appear as well. Plasma frequencies
are eigenfrequencies for charge oscillation, and in a structure with $N_{\rm QW}$
quantum wells there will be $N_{\rm QW}$ such frequencies, with the single
frequency $\Omega $ we have identified corresponding to identical
oscillation in all wells. If the relative phase parameter $\Delta \phi
=2\phi _{\omega }-\phi _{2\omega }$ is constant as the optical pulses
propagate into a sample, only this mode will be excited. \ However, due to
different refractive indices of the sample at $\omega $ and $2\omega $, in
fact $\Delta \phi $ will vary from one well to another, and a number of
plasma eigenmodes will actually be excited. The interesting physics here is
clearly beyond the scope of the present paper.

We also defer to a later communication a discussion of the ultrafast analog
of the intrinsic spin-Hall effect. While we have only treated the
extrinsic Hall effect here, in a (001) GaAs quantum well we can expect an
intrinsic effect if both the Dresselhaus \cite{Dyakonov86} and Rashba \cite{Rashba84} SO
couplings are present, in which case the spin precession rate depends on the
direction of the electron momentum. In such a sample, at typical values of
the Rashba and Dresselhaus couplings the intrinsic-like and extrinsic spin
Hall effects on the photoexcited electrons can be comparable. \ However, in
symmetric quantum wells the analog of the intrinsic spin Hall effect is
absent and, therefore, the kind of experiments we have discussed here will
be sensitive to the extrinsic spin-Hall current.

Such experiments and their description will extend spin-Hall physics beyond
the usual near-equilibrium regime in which it has been studied to date. \
One advantage of the optical experiments we have proposed is that laser
spots can be directed away from any surfaces or edges of a sample, and thus
the treatment of these regions that complicates the analysis of more usual
transport experiments is not present. Another is that, by choosing the
laser photon energy and hence the energy of the injected carriers, the
scattering processes involved can be studied as a function of energy. But
perhaps the main advantage is that ultrafast experiments allow for the study
and isolation of different time scales in the problem. The analysis of this
communication, for example, clearly fails for timescales longer than about a
picosecond, when more complicated space charge dynamics, diffusion, and
recombination effects become important. And indeed even optical experiments
which probe the sample on such longer time scales will require a more
detailed analysis than we present here. But effects on such longer
timescales are irrelevant for the subpicosecond time scale that characterize
the scattering and skew-scattering that are of primary interest. \ And so we
believe it will be time-resolved experiments on the subpicosecond scale, for
which the analysis in this communication provides a starting point, that
will make the most important contribution to our understanding of the
fundamental processes of interest in spintronics.

{\it Aknowledgement} This research was supported by the NSERC and DARPA SpinS program. We are grateful to
I. Rumyantsev and E. Hankiewicz for very valuable discussions.

\newpage

\begin{figure}[tbp]
\includegraphics[width=12.0cm]{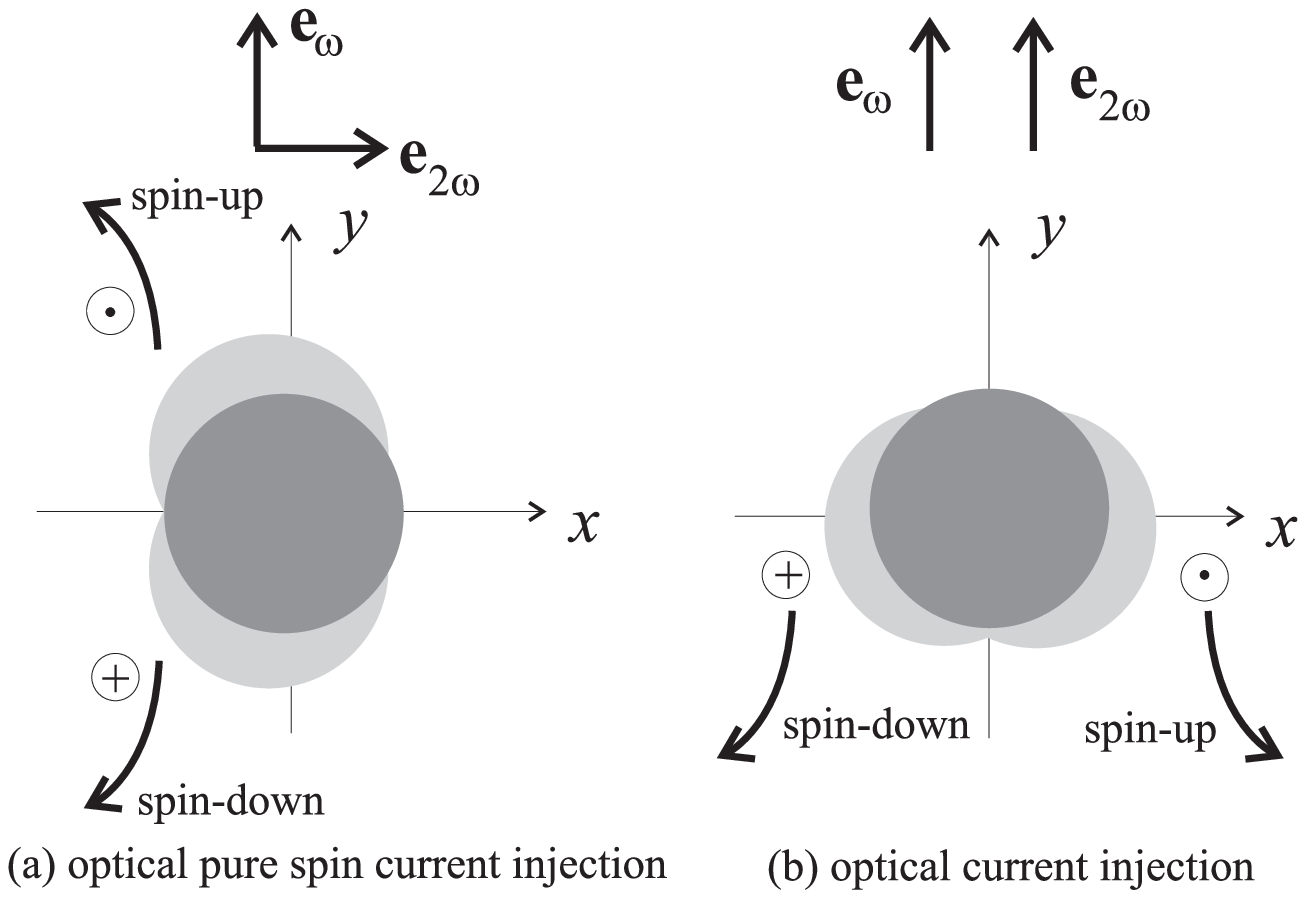}
\caption{ The pure spin current (a) and charge current (b) injection schemes 
for studying the ultrafast spin-Hall effect. The curved arrows show the directions of motion of electrons
with a given spin projection due to the spin-Hall effect. Dark-grey spots
show holes and light-grey spots correspond to the electrons.}
\end{figure}

\newpage

\begin{figure}[tbp]
\vspace{8cm}
\includegraphics[width=8.0cm]{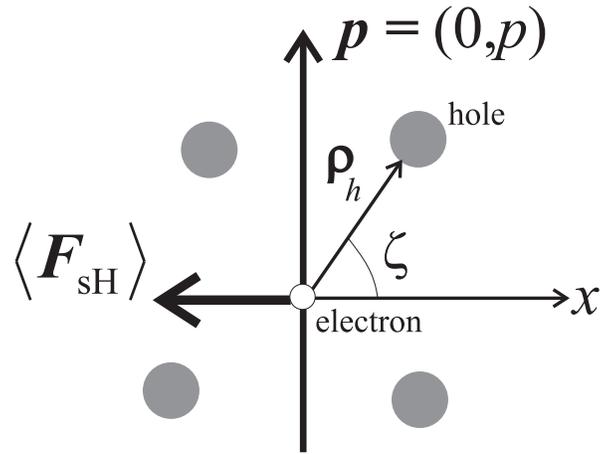}
\caption{ The sketch of an electron with momentum $\mathbf{p}=(0,p)$ interacting with a background 
of holes.}
\end{figure}

\newpage

\begin{figure}[tbp]
\includegraphics[width=8.0cm]{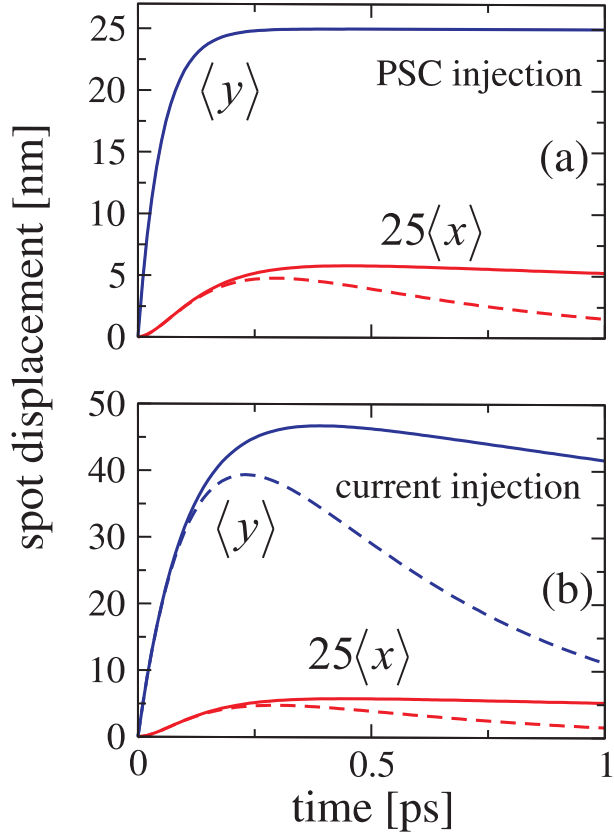}
\caption{ The displacements of spots in the pure spin-current (PSC) (a) and charge current (b)
injection schemes. We adopt (see text) a momentum relaxation
time of $\tau =100$ fs, a spin current relaxation time of $50$ fs, 
a spin Hall time $\tau_{\mathrm{sH}}=10$ ps, and plasma frequencies are 
given by $\Omega \tau =0.15$ (solid lines) 
and $\Omega\tau =0.4$  (dashed lines).  The initial swarm velocity is $500$
km/s. Note that the PSC displacement ($\langle y\rangle$ in Fig. 3a) does not depend on space charge effects
in our model.  }
\end{figure}

\end{document}